\documentclass[pdftex,twocolumn,epjc3]{svjour3}          

\RequirePackage[T1]{fontenc}

\smartqed  

\RequirePackage{graphicx,bm}
\RequirePackage{mathptmx}      
\RequirePackage{flushend}
\usepackage{booktabs}
\usepackage{comment}
\usepackage{rotating}
\RequirePackage[numbers,sort&compress]{natbib}
\RequirePackage[colorlinks,linktoc=all,citecolor=blue,urlcolor=blue,linkcolor=blue]{hyperref}

\usepackage{amsmath,amssymb}
\journalname{Eur. Phys. J. C}

\newcommand{\ce}[1]{Eq.~(\ref{#1})}

\newcommand{\ct}[1]{{Table~\ref{#1}}}

\usepackage[caption=false]{subfig}

\usepackage{calligra}
\usepackage[T1]{fontenc}

\usepackage{blindtext}
\usepackage{verbatim}

\usepackage{subfig}

\usepackage{xspace}

\usepackage{scalerel}
\usepackage{tikz}
\usetikzlibrary{svg.path}

\definecolor{orcidlogocol}{HTML}{A6CE39}
\tikzset{
  orcidlogo/.pic={
    \fill[orcidlogocol] svg{M256,128c0,70.7-57.3,128-128,128C57.3,256,0,198.7,0,128C0,57.3,57.3,0,128,0C198.7,0,256,57.3,256,128z};
    \fill[white] svg{M86.3,186.2H70.9V79.1h15.4v48.4V186.2z}
                 svg{M108.9,79.1h41.6c39.6,0,57,28.3,57,53.6c0,27.5-21.5,53.6-56.8,53.6h-41.8V79.1z M124.3,172.4h24.5c34.9,0,42.9-26.5,42.9-39.7c0-21.5-13.7-39.7-43.7-39.7h-23.7V172.4z}
                 svg{M88.7,56.8c0,5.5-4.5,10.1-10.1,10.1c-5.6,0-10.1-4.6-10.1-10.1c0-5.6,4.5-10.1,10.1-10.1C84.2,46.7,88.7,51.3,88.7,56.8z};
  }
}

\newcommand\orcidicon[1]{\href{https://orcid.org/#1}{\mbox{\scalerel*{
\begin{tikzpicture}[yscale=-1,transform shape]
\pic{orcidlogo};
\end{tikzpicture}
}{|}}}}

\newcommand{\nlo}{\mathrm{NLO}}

\newcommand{\ord}{{\ensuremath{\cal O}}}
\newcommand{\MG}{{\tt MG5aMC}\xspace}
\newcommand{\HwU}{{\tt HwU}\xspace}
\newcommand{\sqrtsnn}{\ensuremath{\sqrt{s_{_{NN}}}}}

\usepackage[normalem]{ulem}

\DeclareMathAlphabet{\pazocal}{OMS}{zplm}{m}{n}

\DeclareFontFamily{OT1}{pzc}{}
\DeclareFontShape{OT1}{pzc}{m}{it}{<-> s * [1.10] pzcmi7t}{}
\DeclareMathAlphabet{\mathpzc}{OT1}{pzc}{m}{it}
\usepackage{txfonts}

\begin{document}

\title{Automated NLO calculations for asymmetric hadron-hadron collisions in {\tt MadGraph5\_aMC@NLO}}

\author{Carlo Flore\thanksref{e1,addr1,addr2,corr}\protect\orcidicon{0000-0002-1071-204X}
\and Daniel Kiko\l{}a\thanksref{e2,addr3}\protect\orcidicon{0000-0001-6896-6475} 
\and Aleksander Kusina\thanksref{e3,addr4}\protect\orcidicon{0000-0002-4090-0084} 
\and Jean-Philippe Lansberg\thanksref{e4,addr5}\protect\orcidicon{0000-0003-2746-5986} 
\and Olivier Mattelaer\thanksref{e5,addr6}\protect\orcidicon{0000-0002-7302-7744}
\and Anton Safronov\thanksref{e6,addr3}\protect\orcidicon{0000-0002-0586-0830}}

\thankstext{e1}{carlo.flore@unica.it}
\thankstext{e2}{daniel.kikola@pw.edu.pl}
\thankstext{e3}{aleksander.kusina@ifj.edu.pl}
\thankstext{e4}{Jean-Philippe.Lansberg@in2p3.fr}
\thankstext{e5}{olivier.mattelaer@uclouvain.be}
\thankstext{e6}{anton.safronov.dokt@pw.edu.pl}

\thankstext{corr}{Corresponding author}

\institute{Dipartimento di Fisica, Università di Cagliari, Cittadella Universitaria, I-09042 Monserrato (CA), Italy \label{addr1} \and
INFN, Sezione di Cagliari, Cittadella Universitaria, I-09042 Monserrato (CA), Italy \label{addr2} \and
Faculty of Physics, Warsaw University of Technology, plac Politechniki 1,00-661, Warszawa, Poland \label{addr3} \and
Institute of Nuclear Physics, Polish Academy of Sciences, ul.~Radzikowskiego 152, 31-342 Cracow, Poland \label{addr4} \and
Université Paris-Saclay, CNRS, IJCLab, 91405 Orsay, France \label{addr5} \and
Centre for Cosmology, Particle Physics and Phenomenology (CP3), Université Catholique de Louvain, Chemin du Cyclotron, Louvain-la-Neuve, B-1348, Belgium \label{addr6}}

\date{\today}

\maketitle

\begin{abstract}
We have extended \texttt{MadGraph5\_aMC@NLO} capabilities by implementing computations for asymmetric hadron-hadron collisions, including proton-nucleus, pion-hadron or nucleus-nucleus collisions in order to obtain a tool for automated perturbative computations of cross sections (or ratios of cross sections) in asymmetric reactions at next-to-leading (NLO) order in $\alpha_S$ in collinear factorisation. This tool, like the original symmetric version of \texttt{MadGraph5\_aMC@NLO}, automatically computes cross sections for different factorisation and renormalisation scales and PDFs provided by the {\tt LHAPDF} library, thereby allowing for an easy assessment of the associated theoretical uncertainties. 
In this paper, we pres\-ent the validation of our code using $W$ and $Z$ boson production in $p$Pb collisions and Drell-Yan-pair production in $\pi W$ collisions. We also illustrate the capabilities of the framework by providing cross section and nuclear modification factors with their uncertainties for the production of $W$ and $Z$ bosons, charm and bottom quarks as well as associated production of $b\bar b+H^0$ in $p$Pb collisions at the LHC.
\end{abstract}

 \tableofcontents

\section{\label{sec:introduction}Introduction}
The main goal of high-energy physics is to study the properties of matter and the interactions of its constituents. Asymmetric hadronic reactions, involving two colliding hadrons of different species ({\it e.g.} protons, mesons and/or nuclei) provide various unique scientific opportunities for studies of different processes which are otherwise difficult to address in symmetric proton-proton or nuclear reactions.

High-energy reactions involving nuclei such as lepton-nucleus ($\ell A$) and proton-nucleus ($pA$) collisions provide valuable insights into nuclear parton distribution functions (nPDFs), which describe the internal structure of nuclei in terms of their constituent quarks and gluons~\cite{Klasen:2023uqj}. nPDFs are an important ingredient of perturbative Quantum Chromodynamics (pQCD) computations within the collinear factorisation framework (see, {\it e.g.}, Ref.~\cite{Collins:1989gx} and Sec.~\ref{sec:formalism}) and are needed for phenomenological studies. They are extracted by fitting experimental data and there are a number of collaborations performing such determinations, including nCTEQ~\cite{Kovarik:2015cma,Duwentaster:2022kpv}, EPPS~\cite{Eskola:2016oht,Eskola:2021nhw}, nNNPDF~\cite{AbdulKhalek:2019mzd,AbdulKhalek:2020yuc,AbdulKhalek:2022fyi}, and others~\cite{Walt:2019slu,Helenius:2021tof,Khanpour:2020zyu,Segarra:2020gtj,Eskola:1998iy,Hirai:2001np,deFlorian:2003qf}.
Several experimental results point to a strong modification of PDFs in a nucleus compared to a free proton~\cite{NewMuon:1996yuf,NewMuon:1996gam,Arnold:1983mw,Gomez:1993ri,BCDMS:1985dor,NuSea:1999egr,NuTeV:2005wsg}. However, so far, our understanding of nPDFs is limited, and specifically unsatisfactory in the case of gluon. This is caused by the fact that data sets used for nPDF fits are either weakly sensitive to the gluon distribution or prompt to other effects obscuring the leading twist factorisation picture, {\it e.g.}~\cite{Vogt:2004dh,Ferreiro:2008wc,Ferreiro:2018wbd,Arleo:2021bpv,Arleo:2020hat}.
An interesting opportunity to improve this situation is provided by heavy-flavour production data from RHIC and the LHC~\cite{Andronic:2015wma,ALICE:2018mml,ALICE:2019fhe,LHCb:2019avm,Kusina:2017gkz,Hadjidakis:2018ifr,Eskola:2019dui,Kusina:2020dki,Chapon:2020heu,Duwentaster:2022kpv,Duwentaster:2021ioo} as well as the EIC~\cite{Boer:2024ylx}.

$pA$ collisions are also important for studies of the matter with partonic degrees of freedom, so-called Quark–Gluon Plasma (QGP). Traditionally, they provide a baseline to identify the effects of ``hot'' nuclear matter ~\cite{Benhar:2023mgk, Elfner:2022iae, Chen:2016dke}. Recently, $pA$ reactions have become interesting on their own in searching for the onset of QGP in small systems. Specifically, a number of experimental results indicate a collective behaviour of particles produced in $pA$ collisions~\cite{ATLAS:2012cix, CMS:2013jlh,  CMS:2015yux, ATLAS:2017hap, ALICE:2018gyx,  ATLAS:2019vcm, CMS:2022bmk, ALICE:2022ruh,  ALICE:2023xiu, ALICE:2023kjg}, and such effects are commonly attributed to the QGP formation.

Asymmetric reactions also offer an interesting opportunity to study the mass number (or system size) dependence of nuclear effects, both related to the QGP formation and so-called cold nuclear matter (like nPDFs)~\cite{ Aduszkiewicz:2015rka, Werner:2018czz,Ojha:2021zbj, Modak:2022jhr, ALICE:2022imr}. While challenging for collider experiments, this kind of reactions can be easily realised with a fixed-target experiment at the Large Hadron Collider (LHC)~\cite{Hadjidakis:2018ifr,Lucarelli:2024ikr}.

Finally, asymmetric reactions provide opportunities beyond nucleus-nucleus collisions of the same type $AA$. For instance, pion-proton ($\pi p$) or pion-nucleus ($\pi A$) reactions can shed light on the poorly known structure of mesons. There are few results from experiments that used pion beams and a nuclear fixed target, such as Drell-Yan production~\cite{Chang:2020rdy} in $\pi^{-}W$, $\pi^{\pm}$Li and $\pi^{-}$Be reactions, and targeted $D^{0}$ and $J/\psi$ production~\cite{McEwen:1982fe, Engel:1992vf, E672:1995won}. These studies provide input for establishing pion PDFs~\cite{Sutton:1991ay,Gluck:1999xe,Wijesooriya:2005ir,Barry:2018ort,Izubuchi:2019lyk,Novikov:2020snp,Cao:2021aci} and test different charm production models. To harvest the full potential of these data, one needs an appropriate tool for theoretical computations for pion-induced collisions.

The interpretation of experimental results very often relies on theoretical computations. There are numerous tools available for such purposes that are able to compute physical observables for many processes within the Standard Model (SM) and Beyond the Standard Model (BSM) both at leading order (LO) and next-to-leading order (NLO) in pQCD, such as \texttt{MadGraph5\_aMC@NLO}~\cite{Alwall:2014hca} (\MG in what follows), \texttt{POWHEG}~\cite{Nason:2004rx,Frixione:2007vw,Alioli:2010xd}, \texttt{MCFM}~\cite{Campbell:2011bn,Campbell:2019dru}, \texttt{HELAC-Onia}~\cite{Shao:2012iz, Shao:2015vga}, \texttt{HERWIG}~\cite{Bellm:2015jjp}, \texttt{PYTHIA}~\cite{Bierlich:2022pfr}, and \texttt{SHERPA}~\cite{Sherpa:2019gpd}. Like most of these codes, \MG is unable to tackle asymmetric hadronic reactions, when two different hadron species collide. Some groups and experimental collaborations maintain modified codes for such cases, but they are usually not publicly available and often not thoroughly tested, while some programs work only at LO. 

Thus, the main motivation of this work is to develop and provide the community with a tool that gives possibilities to perform reliable, precise automated perturbative computations of cross sections at NLO accuracy for SM processes in any asymmetric hadronic reaction, including $pA$, $\pi p$, $\pi A$ or $AB$ collisions. Such computations at NLO in $\alpha_s$ will provide useful theoretical predictions for a wide range of applications, such as the phenomenology of heavy-ion collisions, for the interpretation of the LHC and RHIC data and results of experiments conducted with exotic beams (pions, photons, etc.), providing solid predictions for feasibility studies for new experimental programs.

The paper is organised as follows: Section~\ref{sec:formalism} presents basic elements of collinear factorisation and its implementation in \MG. Then, in Section~\ref{sec:asym-coll}, we introduce modifications needed for asymmetric collisions and provide validation of our implementation. Section~\ref{sec:results} illustrates the capabilities of the tool with a selection of representative computations for asymmetric hadronic processes, while our conclusions are gathered in Section~\ref{sec:conclusions}. Additionally, we provide an appendix where we outline the usage of the new code.

\section{\label{sec:formalism}Formalism}
\subsection{Collinear factorisation framework}
\label{Sec:collinear factorisation framework}
Collinear factorisation~\cite{Bodwin:1984hc,Collins:1985ue,Collins:1998rz,Collins:2011zzd,CTEQ:1993hwr} is the standard and most established approach for calculation of cross sections for high-energy hard processes studied at colliders such as Tevatron, LEP or the LHC. The cross section is calculated as a convolution of a parton-level matrix element, which is process-dependent but calculable within pQCD, and non-perturbative but universal objects, the parton distribution functions (PDFs), which represent the distributions of partons in hadrons. The factorisation theorems have been proven for deep inelastic scattering process as well as for a number of hadronic processes, such as Drell-Yan lepton pair production (DY) in proton-proton ($pp$) collisions~\cite{Ellis:1978ty,Amati:1978by,Collins:1989gx,Bodwin:1984hc,Collins:1985ue,Collins:1988ig}. In the case of an hadronic process such as $hh\to \ell\bar{\ell} X$, the factorisation formula takes the following form: 
\begin{equation}
\begin{aligned}
d\sigma_{hh \to \ell\bar{\ell} X} & = \sum_{a,b}
\int dx_{a} dx_{b} f_{a/h}(x_{a},\mu_{F}) f_{b/h}(x_{b},\mu_{F}) \\
& \times d\widehat{\sigma}_{ab \to \ell\bar{\ell} X}(x_{a},x_{b}, \mu_{F}, \mu_{R})\,
+ \ord\Big(\frac{\Lambda_{\mathrm{QCD}}^2}{\mu_{F}^2}\Big) \, ,
\label{eq:factGEN}
\end{aligned}
\end{equation}
where $f_{a/h}$ and $f_{b/h}$ are the PDFs of the incoming hadrons,  $x_{a}$ and $x_{b}$ are the fractions of the longitudinal momentum carried by the partons compared to the hadrons, $\mu_{F}$ and $\mu_{R}$ are the factorisation and renormalisation scales, and $d\widehat{\sigma}_{ab\rightarrow \ell\bar{\ell} X}$ is the parton-level cross section. Finally, the last term quantifies the approximation made when using this approach which neglects terms that are suppressed as powers of $\Lambda_{\mathrm{QCD}}^2/\mu_{F}^2$.

Factorisation can also be formulated for asymmetric collisions, {\it e.g.}~pion-induced reactions ($\pi p$ or $\pi A$), and reactions involving nuclei (such as $pA$ or $AB$). For these cases the only modification to the factorisation formula of Eq.~\eqref{eq:factGEN} is the change of the PDFs describing the initial state entering the collisions. However, apart from $\pi p$, in most of these cases QCD factorisation has not been formally proven and it is believed that power suppressed terms are enhanced~\cite{Qiu:2003cg,Accardi:2004be}. Moreover, for $AB$ collisions, and partially for $pA$ collisions, it is expected that additional nuclear effects, both from the initial and final states, can play a role~\cite{Gavin:1990gm,Kopeliovich:2001ee,Ferreiro:2008wc,Arleo:2010rb,Sharma:2012dy} and complicate the picture.

When using the collinear-factorisation framework, one needs both PDFs and the parton-level matrix elements. Since the PDFs are non-perturbative but universal objects, they are currently extracted from fits to experimental data. On the other hand, parton-level matrix elements are process-dependent and they need to be calculated process by process at a given order in perturbation theory. Fortunately,  there are currently methods allowing one to automate such calculations for generic processes at both LO and NLO in perturbative QCD, e.g.~\cite{Frixione:1995ms,Catani:1996vz,Gleisberg:2007md,Actis:2016mpe,Cascioli:2011va,Berger:2008sj}. In order to obtain predictions which are precise and accurate enough for a broad class of processes one needs to go at least to NLO. This is what we will do here for asymmetric collisions by adapting \MG\cite{Alwall:2014hca}. 

Before explaining the modifications we have made to \MG to allow for asymmetric collisions, we will briefly recall how NLO calculations are done in \MG. This includes the treatment of real- and virtual-emission contributions as well as subtractions in a way that ensures cancellation of singularities.
Note that we consider here only fixed-order calculations (without parton showering) since collinear factorisation with nuclear beams and parton showering is likely subject to larger factorisation-breaking effects.%
\footnote{The code has not been tested in the more general case when parton showering is matched to the NLO results and, at the moment, should be used in such a configuration only for tests.}

\subsection{NLO computations of symmetric collisions in \MG}
\label{subsec:symColMG5}

We start by recalling the factorisation formula of Eq.~\eqref{eq:factGEN} but showing explicitly where a dependence on quantities related to PDFs appears. Generally, \MG allows one to use either internal hard-coded PDFs, or to access them via the {\tt LHAPDF} library~\cite{Buckley:2014ana}. Our implementation of asymmetric collisions in \MG heavily relies on the second option.
The factorisation formula for the symmetric case, when the collisions happens between two hadrons $h$ of the same species, is given by:
\begin{equation}
\begin{aligned}
d\sigma_{hh \to \ell\bar{\ell} X} & = \sum_{a,b} \int dx_{a} dx_{b} \;
f_{a/h}(x_{a},\mu_{F}; {\tt LHAID\_h}) \\
& \times
f_{b/h}(x_{b},\mu_{F}; {\tt LHAID\_h}) \\
& \times
d\widehat{\sigma}_{ab\to \ell\bar{\ell} X}\left(x_{a},x_{b}, \mu_{F}, \alpha_S(\mu_{R}; {\tt LHAID\_h})\right),
 \label{eq:factSYM}
 \end{aligned}
\end{equation}
where ${\tt LHAID\_h}$ indicates the dependence of the inputs on {\tt LHAPDF}. Note that, in the current implementation of \MG, the parton-level matrix element also depends on the adopted PDFs via the strong coupling $\alpha_S$ which is also retrieved via {\tt LHAPDF}. Formally, $\alpha_S$ should depend only on the renormalisation scale $\mu_R$ and not on the PDFs. However, in practice, one needs to ensure the compatibility of $\alpha_S$ used in the calculation of matrix elements with that used to fit the employed PDFs. This explains the presence of ${\tt LHAID\_h}$ in the parameters of $\alpha_s$ in \ce{eq:factSYM}.

Below we closely follow Ref.~\cite{Frederix:2011ss} to explain how NLO calculations are performed in {\tt MG5}. 
Any cross section at NLO in QCD is decomposed into different types of contributions numbered here by $\alpha$:
\begin{equation}
d\sigma^{\nlo} = \sum_{\alpha} d\sigma^{(\nlo,\alpha)} \,.
\label{eq:decompALPHA}
\end{equation}
The types of possible contributions are $\alpha=\{E, S, C, SC\}$, where $E$ corresponds to \textit{Event}, which stands for a fully-resolved configuration, and $S$, $C$ and $SC$ correspond respectively to the \textit{Soft}, \textit{Collinear},
and \textit{Soft-Collinear} limits. In what follows, $S$, $C$ and $SC$ are referred to as the Count\-er\-e\-vents.%
\footnote{These are closely related to universal subtraction terms for NLO calculation~\cite{Catani:1996eu,Frixione:1995ms,Frixione:1997np}.}
Each of the four $d\sigma^{(\nlo,\alpha)}$ terms in the decomposition of $d\sigma^{\nlo}$ can be written using the factorisation property of Eq.~\eqref{eq:factSYM} as:
\begin{equation}
\begin{aligned}
d& \sigma^{(\nlo,\alpha)} = \\
 & f_{a/h}\big(x_{a}^{(\alpha)},\mu_{F}^{(\alpha)}; {\tt LHAID\_h}\big) \,
f_{b/h}\big(x_{b}^{(\alpha)},\mu_{F}^{(\alpha)}; {\tt LHAID\_h}\big) \\
& \times
W^{(\alpha)}\big(x_{a}^{(\alpha)},x_{b}^{(\alpha)}, \mu_{F}^{(\alpha)}, \alpha_S(\mu_{R}^{(\alpha)}; {\tt LHAID\_h})\big) \;
d\chi_{12} \, d\chi_{n+1}
\label{eq:dsigNLOa}
\end{aligned}
\end{equation}
where the $d\chi_{12}$ is the integration measure over the initial state momentum fractions, and $d{\chi_{n+1}}$ is a measure of $(3n - 1)$ independent variables corresponding to the phase space of the $n+1$ final-state particles.
The quantity $W^{(\alpha)}$ is a weight containing information about the hard matrix elements and can be written as:
\begin{equation}
\begin{aligned}
W^{(\alpha)}  & = g^{2b+2}_{S}\big(\mu_R^{(\alpha)}; {\tt LHAID\_h} \big)
\Biggl[ \widehat{W}_0^{(\alpha)} \\
& + \widehat{W}_F^{(\alpha)} \log\left(\frac{\mu_F^{(\alpha)}}{Q}\right)^2  + \widehat{W}_R^{(\alpha)}
\log\left(\frac{\mu_R^{(\alpha)}}{Q}\right)^2\Biggr] \\
& + g^{2b}_S\big(\mu_R^{(\alpha)}; {\tt LHAID\_h}\big) \; \widehat{W}_{B} \; \delta_{\alpha S}
\label{eq:weight}
\end{aligned}
\end{equation}
where $g_S^2(\mu_R)=4\pi\alpha_S(\mu_{R})$, $b$ is the power of the strong coupling at born level\footnote{This power is not unique for mixed expansion (for instance in $\alpha_s$ and $\alpha$), but the formalism can be easily extended in such case, see \cite{Frederix:2018nkq}. Note that the extension presented in this paper supports such mixed expansion mode.} and $Q$ is the Ellis-Sexton scale which provides a convenient way to parametrise the dependence on the factorisation and renormalisation scales in one-loop computations~\cite{Ellis:1985er}.
Note that the $\widehat{W}^{(\alpha)}$ coefficients also depend on other kinematical variables ({\it e.g.}~the Mandelstam variables); we omit here this dependence for brevity.
The decomposition of $\widehat{W}^{(\alpha)}$ into 
$\widehat{W}_{0}^{(\alpha)}$, 
$\widehat{W}_{F}^{(\alpha)}$,
$\widehat{W}_{R}^{(\alpha)}$,
and $\widehat{W}_{B}$
coefficients is done in such a way that all the dependence on scales is made explicit and the coefficients are scale and PDF independent. Such a decomposition facilitates the possibility to use reweighting procedures in order to vary PDFs or factorisation and/or renormalisation scales. Note that the factorisation and renormalisation scales also carry the $\alpha$ index, meaning that they can be different for different type of contributions (Event, Counterevents).%
\footnote{There is a limitation to this freedom as they need to become the same in the infrared limit. The same holds for momentum fraction-$x$  variables in Eq.~\eqref{eq:dsigNLOa} in the case of event projections, see Appendix A.4 of Ref.~\cite{Frixione:2002ik} for details.}
The $\widehat{W}_{B}$ coefficient originates from the Born level contribution which shares the same kinematics as the Soft Counterevent, and is included in $W^{(S)}$. The  $\widehat{W}_{0}^{(\alpha)}$, $\widehat{W}_{F}^{(\alpha)}$, and $\widehat{W}_{R}^{(\alpha)}$ coefficients fully originate from NLO corrections. The details about their definitions can be found in Appendix~A of Ref.~\cite{Frederix:2011ss} and in Ref.~\cite{Frederix:2009yq}.

While computing cross sections, \MG will keep weight information according to Eq.~\eqref{eq:weight}. This is crucial for an efficient evaluation of scale and PDF uncertainties by reweighting without the need to recompute matrix elements. The $\widehat{W}$ coefficients are calculated only once, and what really need to be recomputed are only the scales, the PDFs and $\alpha_S$.
\footnote{The reweighting factors in the case of NLO computations are no longer simple ratios of PDFs (as in LO) but the same principle of reweighting holds. See Eq.~(2.21) of Ref.~\cite{Frederix:2011ss} for a detailed expression for weights.}
We use this feature of \MG for the extension to asymmetric collisions.

\subsection{Output format of fixed-order calculations in \MG}
\label{subsec:outputMG5}

In the fixed-order (FO) mode of {\tt MG5}, a convenient way to store results of (N)LO calculations is the Histogram with Uncertainty ({\tt HwU}) format.%
    \footnote{For some explanations, see \url{https://answers.launchpad.net/mg5amcnlo/+faq/2671}.}
A {\tt HwU} file contains the histograms of all observables that were requested by the user; they are stored along with possible additional entries containing information about uncertainties of any sorts. By default, every histogram in the {\tt HwU} file contains at least four entries per bin: the bin edges, the cross section value and the (symmetric) uncertainty coming from the MC integration. If the computation of scale and/or PDF uncertainties (triggered via the {\tt reweight\_scale} and {\tt reweight\_PDF} flags in the {\tt run\_card}, {\it i.e.}~the technical configuration file of \MG) are also requested, each histogram will contain additional entries for each bin.
Specifically, for both scale and PDF uncertainties, the information on the central value will be accompanied by the corresponding minimum and maximum values of the observable for each bin ({\it i.e.}~3 entries for both scale and PDF uncertainties per bin). These minimal/maximal values are defined by the envelope of the requested scale variations and the PDF uncertainties are computed via the {\tt LHAPDF} library.
Additionally, cross sections for individual scales used for the variation and for each PDF member are stored. 
In order to count the number of entries per bin in the {\tt HwU} file, we define the number of scale combinations\footnote{The number of scale variations is a product of number of points of the factorisation and renormalisation scales for which the computation is requested: $N_\mu = N_{\mu_F} N_{\mu_R}$.} as $N_{\mu}$ and the number of PDF members as $N_{\rm mem}$. Then, we can write the total number of entries (per bin) in the {\tt HwU} file, assuming that we request both scale and PDF uncertainties to be computed, as:
\begin{equation}
\begin{aligned}
\label{eq:Nsym}
N_{\rm entr} = 
          \underbrace{4}_{\begin{subarray}{l} \text{2 bin edges}\\ \text{1 cross section}\\ \text{1 MC error}\end{subarray}}
        & +\quad
        \underbrace{(3+N_\mu)}_{\begin{subarray}{l} \text{1 central value}\\ \text{2 min/max values}\\ \text{$N_\mu$ scale combinations}\end{subarray}} \\
 &       + \sum_{i=1}^{N_{\rm LHAID}}\underbrace{(3+N_{\rm mem}^{i})}_{\begin{subarray}{l} \text{1 central value}\\ \text{2 min/max values}\\ \text{$N_{\rm mem}$ PDF members}\end{subarray}}.
\end{aligned}
\end{equation}
We have also included here the dependence on $N_{\rm LHAID}$ which accounts for the possibility to perform the calculation for multiple PDFs\footnote{Such an option is already available in the standard \MG code which is able to load multiple {\tt LHAPDF} sets to compute cross sections for symmetric collisions and compute the corresponding uncertainties if the user encodes a list of {\tt LHAIDs} in the {\tt run\_card} file. 
When a calculation for multiple PDFs is requested, the number of PDF members for each {\tt LHAPDF} set can be different. Hence, $N_{\rm mem}^{i}$ depends on the {\tt LHAID} index. Note, however, that in this case the scale variation is computed only for the first {\tt LHAID} from the list.
} at the same time. The $N_{\rm LHAID}$ corresponds to the number of entries for the {\tt LHAID} parameter of the {\tt run\_card}.

\section{\label{sec:asym-coll}Asymmetric collisions in MG5}

\subsection{Implementation}

At NLO, \MG can only handle  symmetric hadron-hadron reactions,%
\footnote{An exception to this are reactions where charge or isospin symmetry can be used (such as $p\overline{p}$ or $pn$).} where the flux for both incoming partons is given by the convolution of two PDFs of the same kind ({\it i.e.}~two proton PDFs or two nuclear PDFs with the same ${\tt LHAID}$). This prevents the use of \MG to obtain predictions for asymmetric reactions involving two different hadronic objects, such as $pA$, $AB$, $\pi p$ or $\pi A$ collisions.
In order to allow for the study of this kind of processes in \MG the existing algorithm implementing Eq.~\eqref{eq:factSYM} needs to be extended. In particular, the possibility to use two distinct PDFs needs to be implemented. At the same time, from Eq.~\eqref{eq:factSYM} it is transparent that one needs to also make certain choice about the $\alpha_S$ which enters through the parton-level matrix element and is related to the adopted PDF set. Below, we write again Eq.~\eqref{eq:factSYM} but for collisions of two different hadrons/nuclei denoted as $h_1$ and $h_2$:
\begin{equation}
\begin{aligned}
d\sigma_{h_1 h_2 \to \ell\bar{\ell} X} & = \sum_{a,b} \int dx_{a} dx_{b} \;
f_{a/h_1}(x_{a},\mu_{F}; {\tt LHAID\_h_1}) \\
& \times f_{b/h_2}(x_{b},\mu_{F}; {\tt LHAID\_h_2}) \\
& \times d\widehat{\sigma}_{ab \to \ell\bar{\ell} X}\left(x_{a},x_{b}, \mu_{F}, \alpha_S(\mu_{R}; {\tt LHAID\_h_2})\right).
 \label{eq:factASYM}
\end{aligned}
\end{equation}
The ${\tt LHAID}$ of the PDF sets that are used are written explicitly indicating our choice of $\alpha_S$ that enters matrix elements. As can be seen from the equation above, the strong coupling is taken from the {\tt LHAPDF} set of particle $h_2$. From the physical point of view it should not matter, as in order to have full compatibility in the calculation, the two PDF sets that are used for the asymmetric collisions should be fully consistent in terms of the used parameters -- in particular they should have the same $\alpha_S$. However, in practice the situation can be more subtle. When doing calculations for asymmetric collisions we need two sets of PDFs for different objects. If we are interested in $pA$ collisions to ensure such compatibility, we can use a given nPDF set together with the proton PDF baseline that was used to extract these nPDFs. On the other hand, when doing calculations for $\pi p$ or $\pi A$ it might not be possible to get fully compatible PDFs (especially in cases when older PDFs are used), as they may have different settings for the $\alpha_S$ running or values of heavy quark masses. For this reason the choice to use $\alpha_S$ of particle $h_2$ can have practical consequences and should be taken into account when using the code. 

While, in the symmetric case, one {\tt LHAID} is sufficient to compute the cross section, in the asymmetric case, Eq.~\eqref{eq:factASYM}, two or more PDF sets with different {\tt LHAIDs} are needed.
In the latter case, within the implementation presented here, one is able to automatically compute different cross sections (together with uncertainties) at once. To trigger the computation for asymmetric collisions, one has only to modify a few parameters in the \MG\ {\tt run\_card}, providing a list of {\tt LHAIDs} (with at least two of them), the first being the baseline (a priori a proton PDF), and the other ones being one or more (n)PDF {\tt LHAPDF} ids which are going to be used for the (n)PDF uncertainties. More details on the usage of the code are given in Appendix \ref{sec:appendix-usage}.

The \HwU output is accordingly extended to account for each PDF-set uncertainty. When performing calculations in the asymmetric mode the following quantities are computed by default: 
\begin{enumerate}
    \item the symmetric cross sections (together with all uncertainties mentioned in Sec.~\ref{subsec:outputMG5}) for each {\tt LHAID} provided;
    \item the asymmetric cross sections for both $h_1h_2$ and $h_2h_1$ collisions where the $h_1$ PDF is taken as the first PDF set in the list of {\tt LHAIDs} and $h_2$ runs over the list of provided {\tt LHAIDs} starting from the second entry.
\end{enumerate}

For the asymmetric collisions, one can request only one {\tt LHAID} for the $h_1$ hadron but multiple {\tt LHAIDs} for the $h_2$ hadron.
When computing the asymmetric $h_1h_2$ and $h_2h_1$ cross sections the PDF uncertainty calculation is always triggered even if the {\tt reweight\_PDF} flag is set to {\tt False}. This will be reflected in the output {\tt HwU} file.%
   \footnote{When PDF uncertainties for asymmetric collisions are requested to be calculated, the number of entries for the asymmetric block in the {\tt HwU} file will contain $2\sum\limits_{i=2}^{N_{\rm LHAID}}\left(3+N^{i}_{\rm mem}\right)$ entries (the factor two is because of $h_1 h_2$ and $h_2 h_1$ combinations, plus 3 for central prediction and maximum/minimum errors, and $N^{i}_{\rm mem}$ due to the number of PDF members). When the PDF uncertainty is not requested there will be still 3 entries for central, maximum and minimum uncertainty values and one entry for single (central) PDF member. In this case all of these entries will be identical.}
For the record the number of entries per bin in an \HwU file for the symmetric and asymmetric collisions are summarised in Table~\ref{tab:HwUcontent}.
Note that, in the modified version of \MG which we present here, we have also modified how scale uncertainties are computed for symmetric collisions. Specifically, in the standard version, when multiple {\tt LHAIDs} are entered in the {\tt run\_card}, the scale uncertainties are computed only for the (central) PDF corresponding to the first {\tt LHAID}. In the modified version, the scale uncertainties are computed for  the (central) PDF of  all the encoded {\tt LHAIDs}. This is reflected in the different numbers of entries in Tab.~\ref{tab:HwUcontent} compared to Eq.~\eqref{eq:Nsym}.\footnote{Although this may vary in some cases, the nPDF sets on {\tt LHAPDF} are by default normalised by the number of nucleons, $A$. Therefore, when computing a cross section for a $pA$ collision, the {\HwU} file will store the normalised cross section $\sigma_{pA}/A$ and not $\sigma_{pA}$.}

Finally, let us mention that since, in asymmetric collisions, the hadronic center-of-mass system (c.m.s.) often differs from the laboratory (Lab) frame, we have allowed for choosing the frame in which results are reported.
In fact, the Lab frame is a natural choice for asymmetric collisions and since, in \MG, the energies of the incoming beams are provided in this frame, we use it as the default one for our asymmetric version of \MG. This means that all the cuts implemented in the {\tt run\_card} are now provided in the Lab frame.
However, if needed, the c.m.s.~frame can be used by setting the {\tt cms\_frame} flag to {\tt True}. This switches both final results and the cuts in the {\tt run\_card} to the c.m.s.~frame (but not the beam energies which are always provided in the {\tt run\_card} in the Lab frame similarly as in the standard version of \MG).
By default, this flag is hidden and has the boolean value {\tt False} - corresponding to the Lab frame.

\begin{table*}[h]
    \centering
    \scalebox{0.95}{
    \begin{tabular}{cccc}
     \hline
       \texttt{reweight\_scale} & \texttt{reweight\_PDF} & No. of entries (symmetric): $N_{\rm entr}^{\rm sym}$ &  No. of entries (asymmetric): $N_{\rm entr}^{\rm asym}$\\
       \hline 
         \texttt{True} & \texttt{True} & 
         $4+ N_{\rm LHAID}\left(3+N_\mu\right) + 
\sum\limits_{i=1}^{N_{\rm LHAID}}\left(3+N_{\rm mem}^{i}\right) $ &         
        $N_{\rm entr}^{\rm sym} + 
2(N_{\rm LHAID}-1)\left(3+N_{\mu}\right) + 
2\sum\limits_{i=2}^{N_{\rm LHAID}}\left(3+N^{i}_{\rm mem}\right)$
         \\
        \texttt{True} & \texttt{False}  & 
          $4+ N_{\rm LHAID}\left(3+N_\mu\right) + 
          N_{\rm LHAID}\left(3+1\right)$
         & 
         $N_{\rm entr}^{\rm sym} + 
2(N_{\rm LHAID}-1)\left(3+N_{\mu}\right) + 
2(N_{\rm LHAID}-1)\left(3+1\right)$
         \\   
         \texttt{False} & \texttt{True}  & 
          $4+\sum\limits_{i=1}^{N_{\rm LHAID}}\left(3+N_{\rm mem}^i\right)$
         & 
         $N_{\rm entr}^{\rm sym} +
        2\sum\limits_{i=2}^{N_{\rm LHAID}}\left(3+N^{i}_{\rm mem}\right)$
         \\
        \texttt{False} & \texttt{False}  & 
          $4 + N_{\rm LHAID}\left(3+1\right)$
         & 
         $N_{\rm entr}^{\rm sym} + 
        2(N_{\rm LHAID}-1)\left(3+1\right)$
         \\
         \hline
    \end{tabular}}
\caption{Number of entries (per bin) in the \HwU output of the modified \MG version for the case of symmetric and asymmetric collisions. Note that, in the symmetric case, the $N_{\rm LHAID}$ is the number of PDF sets for which symmetric collisions will be calculated. Whereas, in the asymmetric case, the $N_{\rm LHAID}$ includes one PDF set for hadron $h_1$ (most often proton) and $(N_{\rm LHAID}-1)$ PDF sets for hadron $h_2$ ({\it e.g.}~different nuclear PDFs). The $h_1$ hadron is assumed to be in the first position in the list of {\tt LHAIDs}.}
    \label{tab:HwUcontent}
\end{table*}

\subsection{Validation}

We now move to the validation of our asymmetric collisions implementation in {\tt MG5}. As there are only a few available NLO computations for asymmetric collisions, we will limit ourselves to two kinds of reactions at different energies: $Z$ and $W^+$ boson production in $p$Pb collisions at the LHC at a c.m.s.~energy per nucleon $\sqrtsnn = 5.02$ TeV, and pion-induced DY pair production at Fermilab at $\sqrtsnn = 21.7$ GeV. 

Let us start with the electroweak boson production at the LHC. We compare our results to those obtained using \texttt{MCFM}~\cite{Campbell:2019dru}\footnote{Note that \texttt{MCFM} by default does not allow for computation of asymmetric collisions. For this purpose grided computations in terms of APPLgrid~\cite{Carli:2010rw} were used, which were previously tested in~\cite{Kusina:2020lyz,Kusina:2016fxy}.} for $Z$ and $W^{+}$ boson production in $p$Pb collisions at the LHC at $\sqrtsnn=5.02$ TeV.
Our computations are performed at NLO accuracy in QCD in the laboratory and c.m.s.~frames, using the following kinematical cuts: $\lvert \eta_{\ell^+}^{\rm lab}\rvert < 2.4$ and $ p_{T}^{\ell^{+}} > 25$ GeV for $\tfrac{d\sigma(W^+ \to \ell^+ \nu_\ell)}{dy^{\ell^+}_{\rm lab}}$, and $\lvert y_{Z}^{\rm cms}\rvert < 3.5$ and $66$ GeV $< m_{\ell^{+} \ell^{-} } < 116$~GeV for $\tfrac{d\sigma(Z\to \ell^+\ell^-)}{dy^Z_{\rm cms}}$.

\begin{figure*}[hbt!]
\centering
\subfloat[]{\label{fig:DY-validation-W}
\includegraphics[width = 6cm, keepaspectratio]{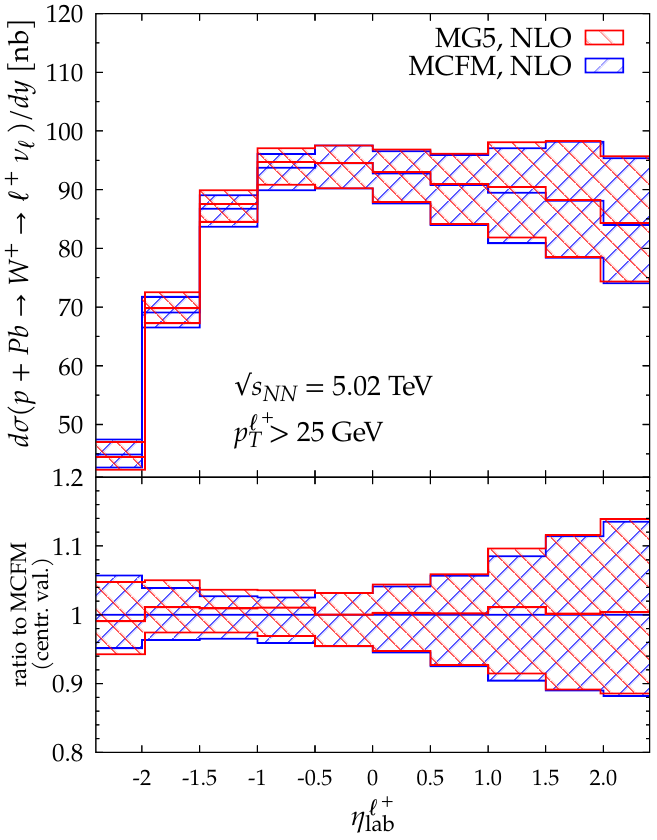}}
\hspace*{5mm}
\subfloat[]{\label{fig:DY-validation-Z}
\includegraphics[width = 6cm, keepaspectratio]{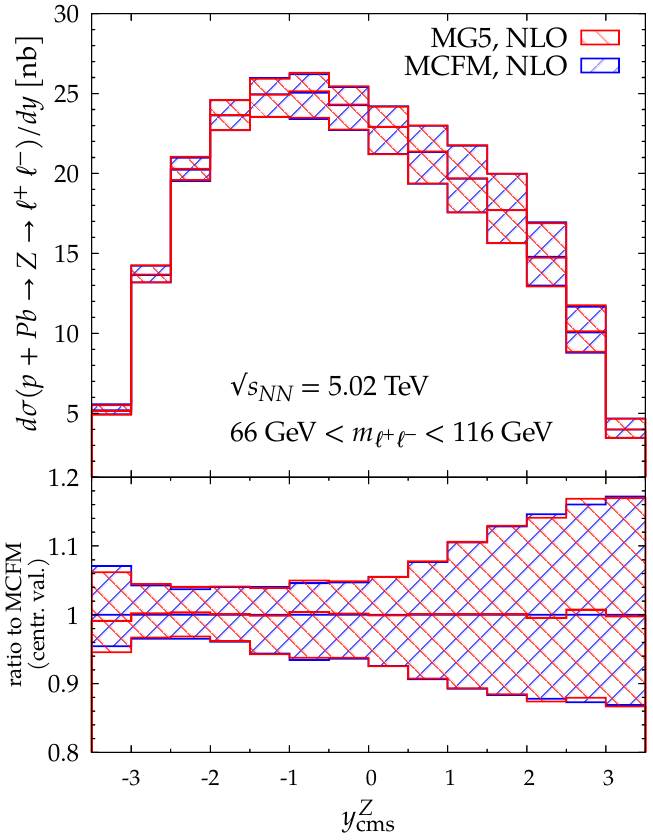}
}
\caption{Comparison of \MG and \texttt{MCFM} results for (a) $W^+$ boson cross-section differential in lepton pseudorapidity, and (b) $Z$ boson cross-section differential in $Z$ rapidity in $p$Pb collisions at $\sqrtsnn=5.02$ TeV at the LHC. Only the nPDF uncertainty is shown.}  
\label{fig:DY-validation} 
\end{figure*}

Fig.~\ref{fig:DY-validation} shows the comparison between results obtained with our modified version of \MG and with \texttt{MCFM}~\cite{Campbell:2019dru}. For our comparison we adopt the CT10nlo proton PDFs~\cite{Lai:2010vv} and the nCTEQ15 nuclear PDFs~\cite{Kovarik:2015cma}. In the upper panels we show the comparison of the cross section together with the associated nPDF uncertainty. 
The red histograms show the results for \MG and the blue ones for \texttt{MCFM}. 
In the bottom panels of Fig.~\ref{fig:DY-validation} we present the ratio of the results from the upper panels normalised to the central value of {\tt MCFM}. All plots show excellent agreement between our new implementation of asymmetric collisions in \MG and the results obtained with \texttt{MCFM}, both for the cross section and the associated nPDF uncertainty.

To further validate our implementation, we consider the DY process $\pi^{-} W \to \mu^{+}\mu^{-}X$ in the hadronic c.m.s.~frame within the kinematics of the E615 experiment~\cite{Conway:1989fs}, using a 252 GeV pion beam on a tungsten target,  resulting in $\sqrt{s} = 21.7$ GeV.
The JAM collaboration used these data in their determination of the $\pi$ PDFs~\cite{Cao:2021aci}. We have compared our NLO computation with their prediction, given as a ratio to the E615 data.
Given the low energy, the process covers kinematical region of low invariant masses. Thus, it is mediated by a virtual photon, and the $Z$ channel was neglected in the JAM analysis. Hence we make the same approximation.\footnote{Including the contributions via $Z$ bosons (together with interferences with photon) is however straightforward in \MG at the level of the process generation.} 
The invariant-mass region of the produced muons measured by E615 is $4.05$ GeV $< m_{\mu\mu} < 8.55$~GeV, and the cross section is given as a function of the Feynman-$x$ variable $x_{F}$, defined as:
\begin{equation}
 x_{F}=2\sqrt{\tau}\sinh{y_{\mu\mu}},
\end{equation}
where $\tau=m_{\mu\mu}^{2}/s$, $y_{\mu\mu}$ is the rapidity of the muon pair, and $s$ is the square of the c.m.s.~energy of the collision.

\begin{figure*}[hbt!]
\centering
\includegraphics[scale=0.75]{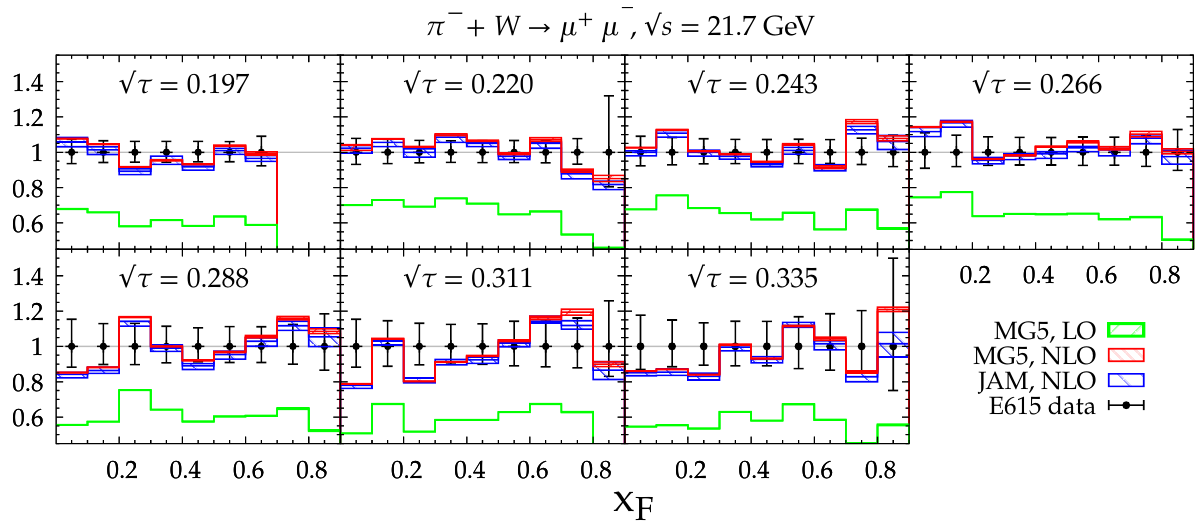}
\put(-465,35)
{
  \makebox(20,20)[lb]
  {
\Large 

\begin{sideways}
\textit{
${
{\frac{ {d^{2} \sigma}^{({\rm MG5/JAM})} }  { d\sqrt{\tau} dx_{F} }} / {\frac{ {d^{2} \sigma}^{({\rm E615})} }  { d\sqrt{\tau} dx_{F} }}
}$  
}
\end{sideways}
  }
}
\caption{The NLO (red) and LO (green) \MG cross sections and those obtained by JAM at NLO (blue) (Fig.~5 (top) of Ref.~\cite{Cao:2021aci}) normalised to the E615 experimental data for the $\pi^{-} W \to \mu^{+}\mu^{-}X$ Drell-Yan cross section as a function of $x_F$ in different $\sqrt{\tau}$ bins. The Monte Carlo uncertainties of the \MG and JAM calculations are shown.}
\label{fig:piW-validation} 
\end{figure*}

We compare our \MG-based computation for $\tfrac{d^2\sigma}{d\sqrt{\tau}dx_F}$ with with that of the JAM Collaboration~\cite{Cao:2021aci} in different $\sqrt\tau$ bins as a function of $x_F$. Both predictions are computed adopting the JAM21 pion PDFs~\cite{Cao:2021aci} and the EPPS16 nPDFs~\cite{Eskola:2016oht}\footnote{In the case of the JAM21 pion PDFs, which are provided as a MC set of replicas, we use the average replica and for EPPS16 the central PDF and we do not compute any PDF uncertainty for this validation.}. Fig.~\ref{fig:piW-validation} shows the ratio of our computations and the JAM values to the experimental data.
The \MG and JAM error bands represent the MC statistical uncertainties which tend to grow in the most forward region where one of the parton-momentum fractions approaches unity. The experimental uncertainties are shown in black. 
\MG and JAM NLO results are essentially identical and are compatible with the experimental results within uncertainties while the LO \MG results are significantly smaller than the data.

\section{\label{sec:results}Illustrative asymmetric computations}
\renewcommand{\arraystretch}{1.2}
\setlength{\tabcolsep}{0.5em}

We can now move to a selection of computations made with the asymmetric version of \MG. 
We will show not only cross sections, but also nuclear modification factors. 
The idea behind this section is to demonstrate the new capabilities of the extended \MG code (in particular as what regards the automatic evaluation of uncertainties)  but not to perform phenomenological studies. Our comparisons to data are purely illustrative and it is not our intention here to assess how well a given nPDF accounts for the observed nuclear effects. As a matter of fact, we will only show results for the nCTEQ15 and EPPS16 sets and not more recent nPDFs, like~\cite{Duwentaster:2022kpv,Eskola:2021nhw,AbdulKhalek:2022fyi,Helenius:2021tof,Khanpour:2020zyu}.

\subsection{Vector-boson production}\label{sec:res_xsec}

We start by showing \MG results for $Z$ and $W^{\pm}$ boson production in $p$Pb collisions along with the experimental data collected by the ATLAS~\cite{ATLAS:2015mwq} and CMS collaborations~\cite{CMS:2015zlj,CMS:2015ehw}. 
Tab.~\ref{tab:Z-W-kinematics} lists processes and kinematical cuts applied to the bosons and leptons.

\begin{table*}[htb!]
\centering
\begin{tabular}{ p{3cm} p{3.5cm} p{8.5cm} }
 \hline
 Experiment & Observable & Kinematical cuts\\
 \hline
 ATLAS~\cite{ATLAS:2015mwq} & $d\sigma(Z \to \ell^+ \ell^-) / dy^{Z}_{\rm cms}$  & $\lvert y^{Z}_{\rm cms}\rvert < 3.5$, $66$ GeV $< m_{\ell^+ \ell^-} < 116$ GeV\\
 CMS~\cite{CMS:2015zlj}  & $d\sigma(Z \to \ell^+ \ell^-) / dy^{Z}_{\rm cms}$  & $|\eta^{\ell^{\pm}}_{\rm lab}| < 2.4$, $ p_{T}^{\ell^+ (\ell^-) }  > 20$ GeV, $66$ GeV $< m_{\ell^+ \ell^- } < 116$ GeV\\
 CMS~\cite{CMS:2015ehw}  & $d\sigma(W^+ \to \ell^+ \nu) / d\eta^{\ell^+}_{\rm lab}$ & $\lvert \eta_{\rm lab}^{\ell^+} \rvert < 2.4$, $ p_{T}^{\ell^+} > 25$ GeV\\
 CMS~\cite{CMS:2015ehw}  & $d\sigma(W^- \to \ell^-\bar{\nu}) / d\eta^{\ell^-}_{\rm lab}$ & $\lvert \eta_{\rm lab}^{\ell^-}\rvert < 2.4$, $p_{T}^{\ell^{-}} > 25$ GeV \\
 \hline
\end{tabular}
\caption{Summary of kinematical cuts adopted by ATLAS and CMS measurements used in Figs.~\ref{fig:pPbtoZ} and \ref{fig:pPbtoWpm}.}
\label{tab:Z-W-kinematics}
\end{table*}

\begin{figure*}[ht!]
\begin{center}
\subfloat[nCTEQ15 \& ATLAS]{\label{fig:pPbtoZ_ATLAS_nCTEQ15}
\includegraphics[width = 6cm, keepaspectratio]{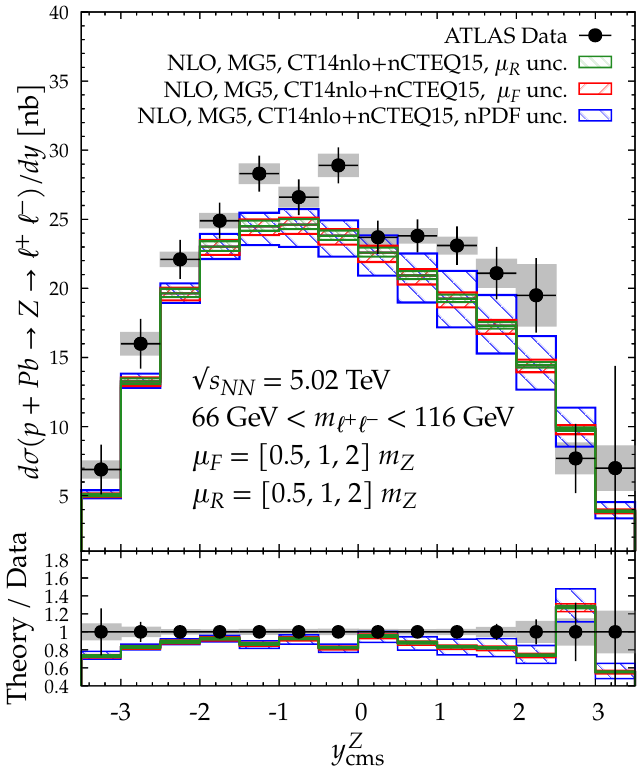}}
\hspace{5mm}
\subfloat[EPPS16 \& ATLAS]{\label{fig:pPbtoZ_ATLAS_EPPS16}
\includegraphics[width = 6cm, keepaspectratio]{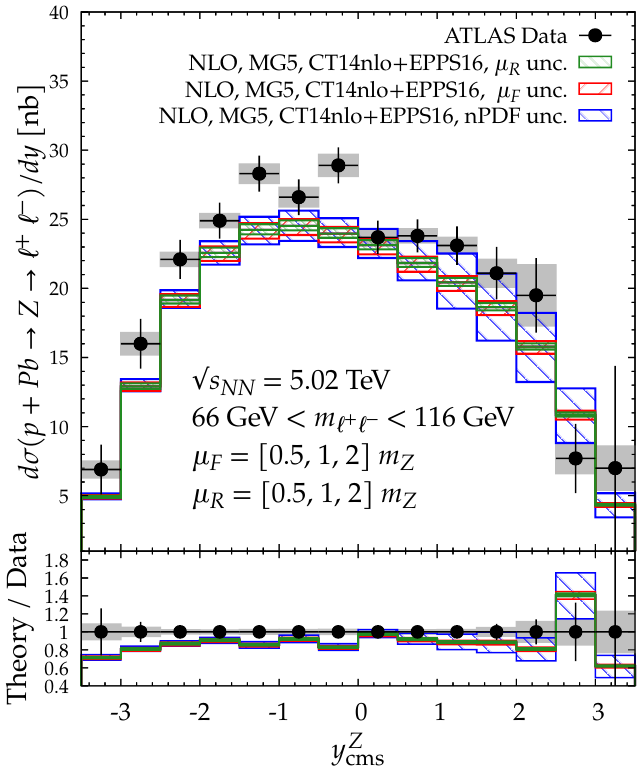}}
\\
\subfloat[nCTEQ15 \& CMS]{\label{fig:pPbtoZ_CMS_nCTEQ15}
\includegraphics[width = 6cm, keepaspectratio]{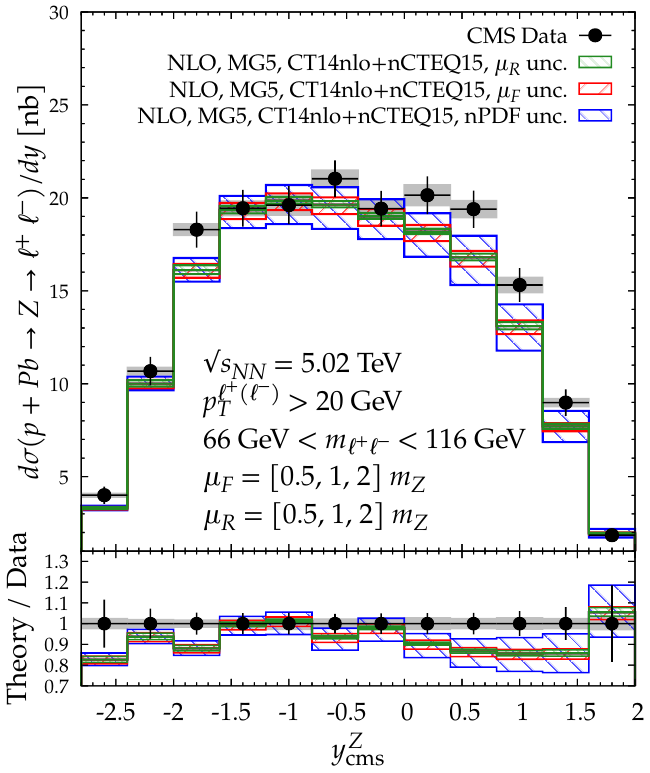}}
\hspace{5mm}
\subfloat[EPPS16 \& CMS]{\label{fig:pPbtoZ_CMS_EPPS16}
\includegraphics[width = 6cm, keepaspectratio]{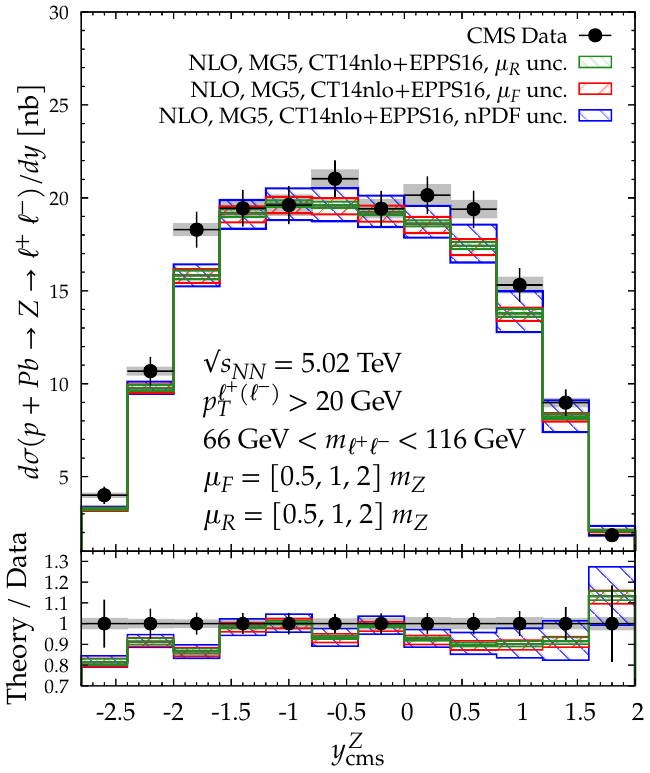}}
\caption{
NLO \MG computation of the rapidity differential cross section for the $Z$ boson production in $p$Pb collisions at $\sqrtsnn=5.02$ TeV using the CT14nlo+nCTEQ15 (a,c) and CT14nlo+EPPS16 (b,d) (n)PDFs, compared with ATLAS~\cite{ATLAS:2015mwq} (a,b) and CMS~\cite{CMS:2015zlj} (c,d) experimental data. Scale and nPDF uncertainties are automatically computed by \MG.
} 
\label{fig:pPbtoZ}
\end{center}
\end{figure*}

\begin{figure*}[h]
\centering
\subfloat[nCTEQ15 for $W^+$]{\label{fig:pPbtoWpm_1}
\includegraphics[width=6cm]{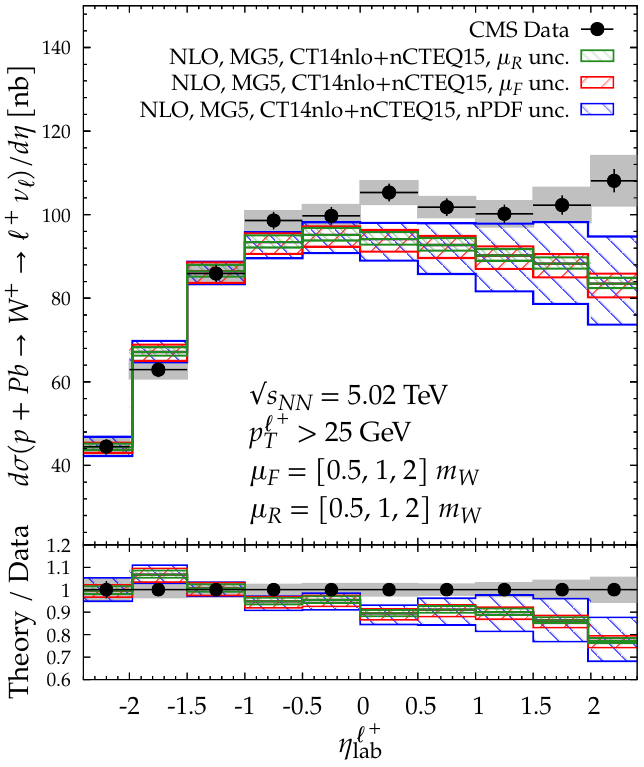}}
\hspace{5mm}
\subfloat[EPPS16 for $W^+$]{\label{fig:pPbtoWpm_2}
\includegraphics[width=6cm]{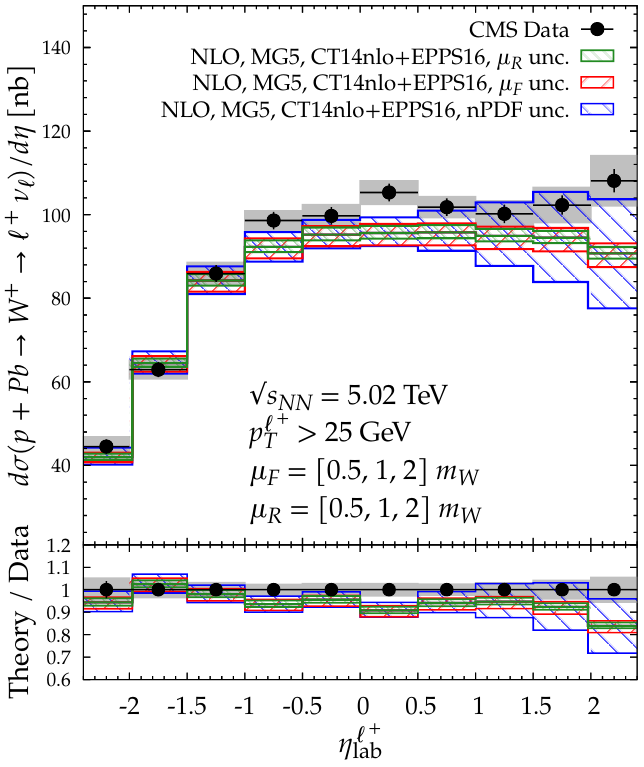}}
\\
\subfloat[nCTEQ15 for $W^-$]{\label{fig:pPbtoWpm_3}
\includegraphics[width=6cm]{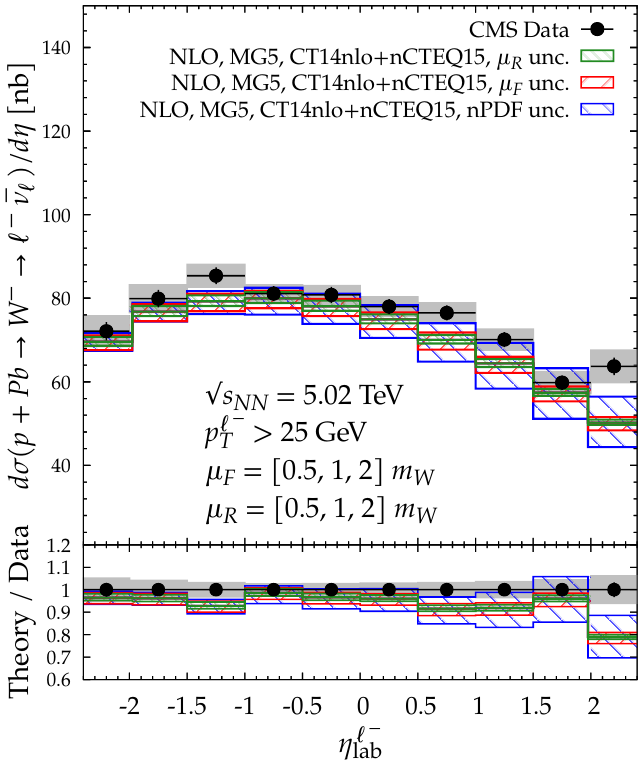}}
\hspace{5mm}
\subfloat[EPPS16 for $W^-$]{\label{fig:pPbtoWpm_4}
\includegraphics[width=6cm]{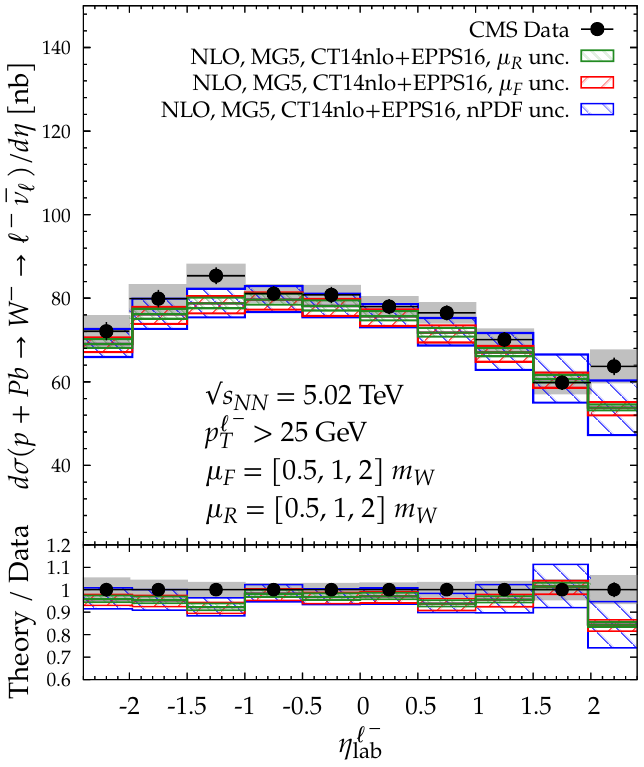}}
\caption{
NLO \MG computation of the lepton laboratory-pseudorapidity differential cross section for $W^+$ (a,b) and $W^-$ (c,d) boson production  in $p$Pb collisions at $\sqrtsnn=5.02$ TeV using the CT14nlo+nCTEQ15 (a,c) and CT14nlo+EPPS16 (b,d) (n)PDFs compared with the CMS~\cite{CMS:2015ehw} experimental data. Scale and nPDF uncertainties are automatically computed by \MG. 
} 
\label{fig:pPbtoWpm} 
\end{figure*}

Fig.~\ref{fig:pPbtoZ} shows results for the $y^Z_{\rm cms}$-differential cross sections for $Z$-boson production in $p$Pb collisions at $\sqrt{s} =5.02$ TeV for the cuts of \ct{tab:Z-W-kinematics} and experimental results from ATLAS~\cite{ATLAS:2015mwq} (Figs.~\ref{fig:pPbtoZ_ATLAS_nCTEQ15} and \ref{fig:pPbtoZ_ATLAS_EPPS16}) and CMS~\cite{CMS:2015zlj} (Figs.~\ref{fig:pPbtoZ_CMS_nCTEQ15} and \ref{fig:pPbtoZ_CMS_EPPS16}). We adopted the CT14NLO PDFs~\cite{Dulat:2015mca} for the proton, and EPPS16nlo and nCTEQ15 nPDFs for the lead nucleus. The ratio of the \MG results to the data is shown in the lower panels of each plot. The nPDF uncertainty is shown with blue bands, while the $\mu_F$ and $\mu_R$ scale variations are separately shown with red and green bands, respectively. Using the same settings for the proton and nuclear PDFs, Fig.~\ref{fig:pPbtoWpm} compares the \MG results for $W^+$ and $W^-$ production in $p$Pb collisions with CMS measurements~\cite{CMS:2015ehw}. Note again that, thanks to the \MG extension presented here, such plots can be obtained in a single run by indicating three {\tt LHAIDs} in the {\tt run\_card}. In both cases, the dominant uncertainty at forward (pseudo)rapidity is the nPDF one, while the size of scale and nPDF uncertainties becomes comparable in the backward rapidity region. This information is now automatically computed. Additional results for any nPDF would also be obtained straightforwardly by adding their {\tt LHAIDs} in the {\tt run\_card}, but we recall that it is not our purpose here to make a survey of data-theory comparisons.

\subsection{Nuclear modification factors}\label{sec:RpA}

When considering collisions including nuclei (either $pA$ or $AB$), Nuclear Modification Factors (NMFs) are usually derived and compared to data. This is partly driven by the possibility of cancellation of different uncertainties in ratios and partly by a more explicit demonstration of deviations from certain baseline observables. To make such calculations easier, we provide the possibility to directly compute NMFs in our asymmetric extension of {\tt MG5}. 
Below we provide some more details on how to compute them and show some illustrative results for NMFs.

\begin{figure*}[h!]
\centering
\subfloat[]{\label{fig:R_pPB_ALICE_a}
\includegraphics[width=6.5cm, keepaspectratio]{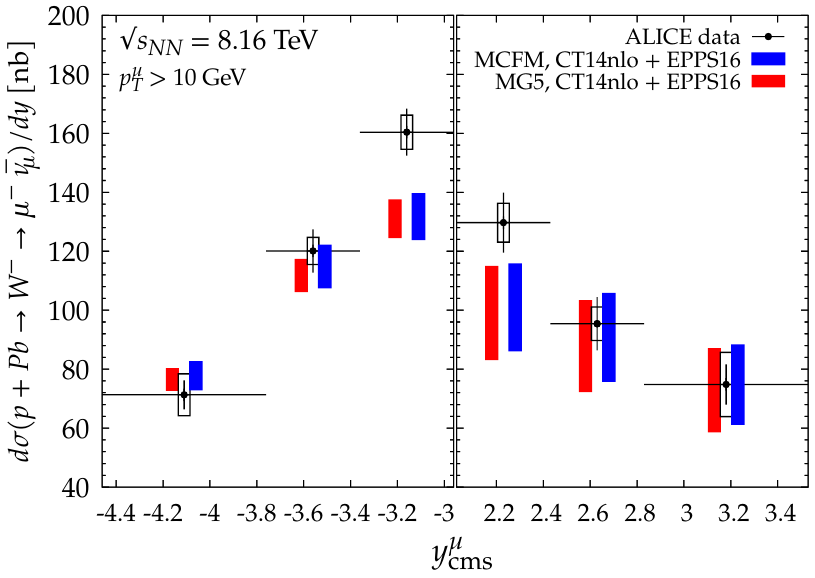}
}
\hspace{3mm}
\subfloat[]{\label{fig:R_pPB_ALICE_b}
\includegraphics[width=6.5cm, keepaspectratio]{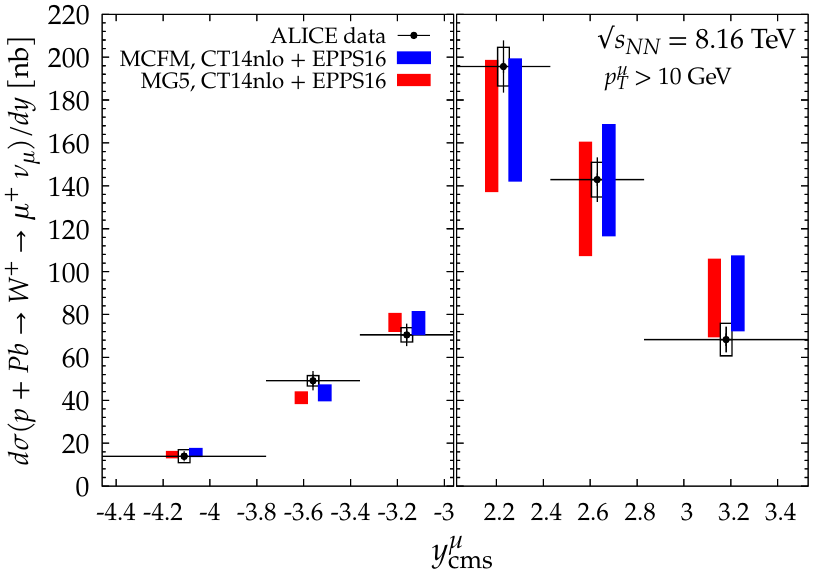}
}
\\
\subfloat[]{\label{fig:R_pPB_ALICE_c}
\includegraphics[width=6.5cm, keepaspectratio]{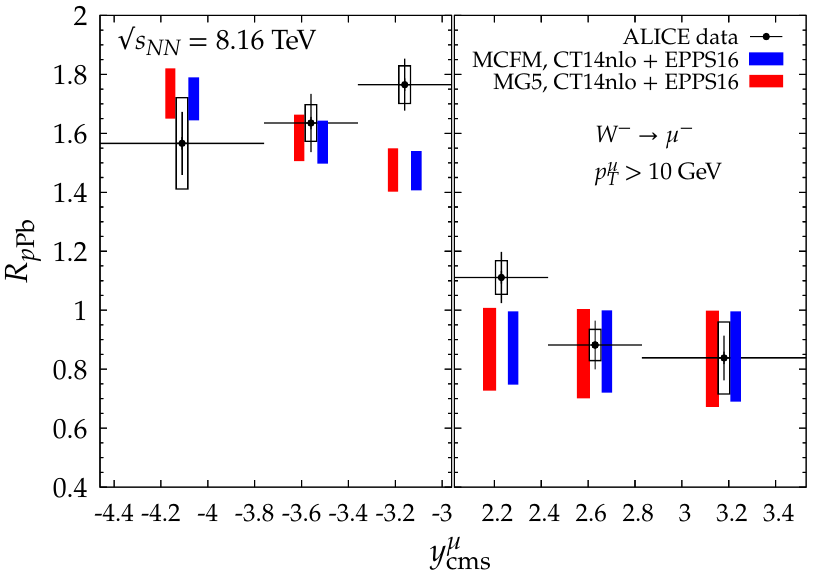}
}
\hspace{3mm}
\subfloat[]{\label{fig:R_pPB_ALICE_d}
\includegraphics[width=6.5cm, keepaspectratio]{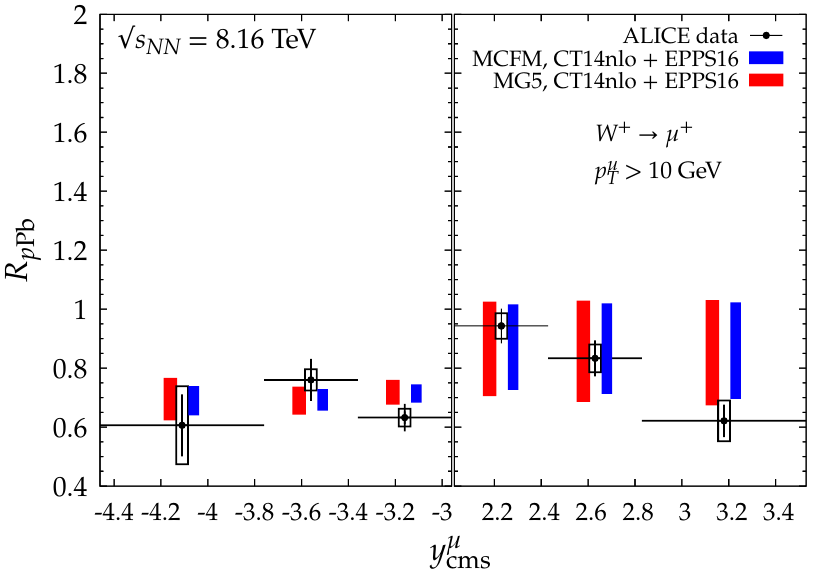}
}
\caption {
Comparison of \MG and \texttt{MCFM} results and ALICE data~\cite{ALICE:2022cxs} at $\sqrtsnn = 8.16$ TeV  for the $y^\mu_\text{cms}$ differential cross sections (a, b) and $R_{p{\rm Pb}}$ (c, d) as a function of $y^\mu_\text{cms}$ for $W^-$ (a, c) and $W^+$ (b, d) production in $p$Pb collisions (with $p_{T}^{\mu} > 10$ GeV).} 
\label{fig:R_pPb_ALICE} 
\end{figure*}

For a generic $AB$ collision involving two nuclei $A$ and $B$ with atomic numbers $A_A$ and $A_B$ respectively, the NMF reads
\begin{equation}
R_{AB} \equiv \frac{1}{A_A A_B} \frac{d\sigma_{AB}}{d\sigma_{pp}},
\end{equation} 
where $\sigma_{AB}$ and $\sigma_{pp}$ are respectively the cross sections for  $AB$ and proton-proton collisions. In the case of proton-lead collisions, the nuclear modification factor simplifies to:
\begin{equation}
R_{p \rm Pb} \equiv \frac{1}{A_{\rm Pb}} \frac{d\sigma_{p{\rm Pb}}}{d\sigma_{pp}}.
\end{equation}

Such a ratio is usually adopted as a proxy for nPDFs, as some theoretical and experimental uncertainties are expected to cancel to some extent within the ratio. It is usually measured as a function of various kinematical variables. The extension of \MG to asymmetric collisions provides an automated tool that allows one to compute in a single run both cross sections and nuclear modification factors. Again, all the relevant uncertainties ({\it e.g.}~scale and (n)PDF ones) are automatically computed. In the current implementation, we adopt an approximation to propagate the nPDF uncertainty on $R_{AB}$. In fact, at least in the kinematical regions currently probed by different experiments, the uncertainty on proton PDFs is usually negligible compared to the one for nPDFs. So, to automatically compute $R_{AB}$ in a single \MG run, we normalise the nPDF uncertainty to the central value of the baseline $pp$ cross section as
\begin{equation}
\delta(R_{AB}) =  \frac{1}{A_A A_B}   \frac {\delta(d\sigma_{AB})}{d\sigma_{pp}^{\rm central}}\,.
\end{equation}

It must be mentioned that the information about nuclear modification factors (values, uncertainties, etc.) is not directly stored in the HwU file. The corresponding NMF information, as a function of kinematic variables, is generated at the plotting level using cross sections that are already in the HwU file.

\begin{figure}[h!]
\subfloat[]{\label{fig:pPbtoccbar}
\includegraphics[width=0.8\columnwidth,keepaspectratio]{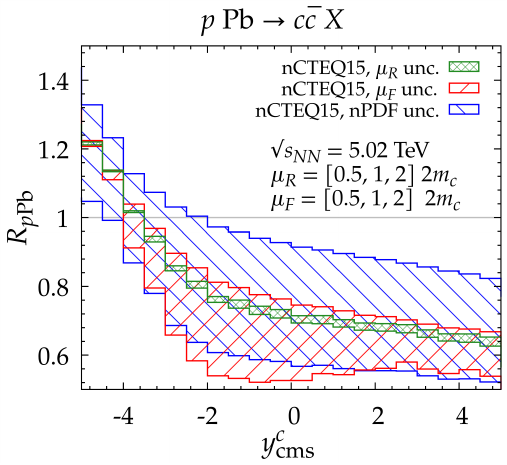}
}\\
\subfloat[]{\label{fig:pPbtobbbar}
\includegraphics[width=0.8\columnwidth,keepaspectratio]{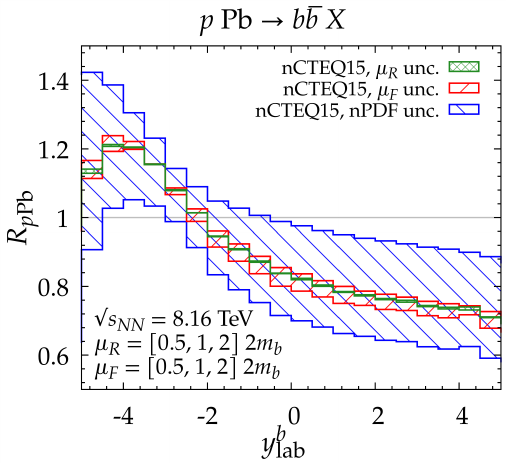}
}\\
\subfloat[]{\label{fig:pPbtoHbbbar}
\includegraphics[width=0.8\columnwidth,keepaspectratio]{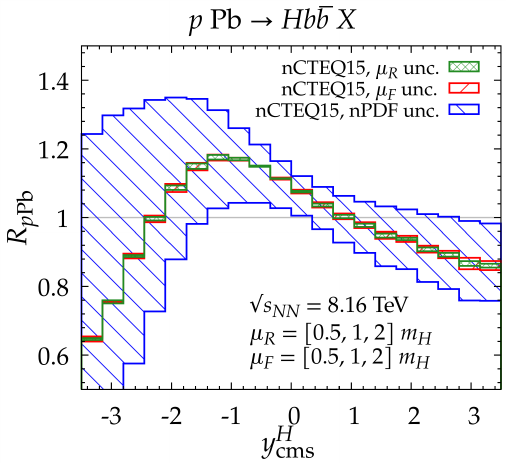}
}
\caption {$R_{p{\rm Pb}}$ predictions for $c\bar c$, $b \bar b$, $H + b\bar b$ inclusive production in $p$Pb collisions at $\sqrtsnn=5.02$ TeV and $\sqrtsnn=8.16$ TeV as a function of respectively the rapidity of the $c$ quark, $b$ quark and $H$ boson. The nPDF, $\mu_F$ and $\mu_R$ uncertainties are shown with blue, red and green bands, respectively.}
\label{fig:R_pPb_predictions} 
\end{figure}

In Fig.~\ref{fig:R_pPb_ALICE} we compare our {\MG}-based predictions for $W^\pm$ production in $p$Pb collisions at $\sqrtsnn =8.16$ TeV with the ones based on {\tt MCFM} and against the experimental data measured by ALICE~\cite{ALICE:2022cxs} for both cross section and $R_{p{\rm Pb}}$. These observables are presented as a function of the c.m.s.~rapidity of the decay muon in the backward ($-4.46 < y_{\rm cms}^{\mu} < -2.96$)  and forward ($2.03 < y_{\rm cms}^{\mu} < 3.53$) regions.  
The computations are made with the CT14NLO PDFs~\cite{Dulat:2015mca} for the proton and the EPPS16~\cite{Eskola:2016oht} nPDFs for the lead nucleus. The uncertainty bands represent here the nPDF uncertainty (the scale uncertainties are not shown). The slight difference in the uncertainty bands for \MG and {\tt MCFM} can be explained by the different computation methods of the uncertainties used in the two calculations: the former computes asymmetric uncertainty as advocated by the used nPDF sets while the latter used symmetric ones. Again, we find a good agreement between both codes, keeping in mind that the asymmetric version of MCFM is not public and not automated.

We conclude here with some illustrative computations for potential future $R_{p{\rm Pb}}$ measurements in the LHC kinematics. As \MG is a fully automated code, now integrating the possibility to compute observables for asymmetric hadronic reactions, it is indeed possible to provide predictions for any hard process in $pA$ or $AB$ reactions. 
In Fig.~\ref{fig:R_pPb_predictions}, we present $p_T$-integrated $R_{p{\rm Pb}}$ computations for $c\bar c$ production at $\sqrtsnn = 5.02$ TeV (Fig.~\ref{fig:pPbtoccbar}), and for $b\bar b$ (Fig.~\ref{fig:pPbtobbbar}) and associated Higgs-$b\bar b$ production (Fig.~\ref{fig:pPbtoHbbbar}) at $\sqrtsnn = 8.16$ TeV as a function of respectively the rapidity of the $c$ quark, $b$ quark and $H$ boson. The predictions are computed using the nCTEQ15 nPDFs with corresponding proton baseline. The nPDF uncertainty is shown with blue bands, while the $\mu_F$ and $\mu_R$ scale uncertainties are respectively shown with red and green bands. Note that Figs.~\ref{fig:pPbtoccbar} and~\ref{fig:pPbtoHbbbar} are plotted using {\tt cms\_frame = True}, hence setting both beam energies equal to $\sqrtsnn/2$ in the {\tt run\_card} with the kinematical variables defined in the c.m.s.~frame. Fig.~\ref{fig:pPbtobbbar} is generated in the laboratory frame with {\tt cms\_frame = False} and using the actual beam energies, 6.5~TeV and 2.56~TeV, in the {\tt run\_card} with the kinematical variables defined in the laboratory frame.

\section{\label{sec:conclusions}Conclusions}
We have presented an extension of the \MG framework to asymmetric hadronic collisions, such as $pA$, $AB$, $\pi p$ and $\pi A$, allowing one to perform any asymmetric calculations at next-to-leading order in $\alpha_s$ for any SM and BSM process.
The validation of the code was done using $W$ and $Z$ bosons production in $p$Pb collisions at the LHC, and Drell-Yan pair production in $\pi$W reactions. All tests show good agreement between the published results and the result obtained with our new implementation. As a demonstration of the tool capabilities, we have also presented examples of cross section calculations at NLO in QCD for $c$- and $b$-quark production, as well as for associated Higgs + $b\bar b$ production, in $p$Pb collision at the LHC kinematics.

We believe that the presented code will be particularly useful for the heavy-ion community providing new possibilities for phenomenological studies of heavy-ion collisions but also more generally allowing to study any hadron-hadron collisions. Such a tool was missing for a long time and we are confident it will be used already for the analyses of Run 3 data from the LHC.

The code is publicly available for download at \url{https://nloaccess.in2p3.fr} and will be made available for online usage at \url{https://nloaccess.in2p3.fr/tools/MG5/index} un\-der the NLOaccess Virtual Access~\cite{Flore:2023dps}, where users are free to test it or to work with it and the other codes available on the online platform.

\section*{Acknowledgements}
We thank P. Barry, V. Bertone, R. Frederix, C. Flett,  A.C. John Rubesh Rajan, K. Lynch, I. Schienbein and H.S. Shao, for useful discussions.

This project has received funding from the European Union’s Horizon 2020 research and innovation programme under the grant agreement No.824093 (STRONG-2020) in order to contribute to the EU Virtual Access {\sc NLOAccess}.
This project has also received funding from the French ANR under the grant ANR-20-CE31-0015 (``PrecisOnium'').
A.K. acknowledges the support of Narodowe Centrum Nauki under Sonata Bis Grant No. 2019/34/E/ST2/00186.
C.F.~is supported by the European Union ''Next Generation EU" program through the Italian PRIN 2022 grant n. 20225ZHA7W.
OM acknowledges support by FRS-FNRS (Belgian National Scientific Research Fund) IISN projects 4.4503.16 (MaxLHC) which directly support the full \MG project.
A.S and D.K. declare that their research was funded by POB HEP of Warsaw University of Technology within the Excellence Initiative: Research University (IDUB) programme within the IDUB-POB-FWEiTE-2 project grant.
This work was also partly supported by the French CNRS via the IN2P3 projects GLUE@NLO and QCDFactorisationAtNLO as well as via the COPIN-IN2P3 project \#12-147 “kT factorisation and quarkonium production in
the LHC era”, by the Paris-Saclay U. via the P2I Department and by the GLUODYNAMICS project funded by the "P2IO LabEx (ANR-10-LABX-0038)" in the framework "Investissements d’Avenir" (ANR-11-IDEX-0003-01) managed by the Agence Nationale de la Recherche (ANR), France.

\appendix
\section*{\label{sec:appendix-usage}Appendix: example of code usage}

In this appendix, we give more details about how to run our updated version of \MG in the asymmetric mode. We emphasise that this implementation works only in the NLO QCD fixed-order mode, and through {\tt LHAPDF}. As an example, we consider the $Z$ boson production decaying in dielectrons. The process generation syntax is the same as for the default implementation for $pp$ collisions:

\begin{verbatim}
generate p p > Z > e+ e- [QCD]
\end{verbatim}

After the code generation stage, the {\tt run\_card} has to be modified. In this version, the {\tt run\_card} contains the following new commented lines:
\begin{verbatim}
#********************************************
# Choice of the collision system in terms 
#    of the PDFs:
#
#    True / False = asymm_choice, 
#       where:
#    False means symmetric 
#    True  means asymmetric
#       
# In the "symmetric" case the default 
# MG5aMC_NLO calculation will be used.
# In the "asymmetric" case, the first 
# PDF in the list of LHAIDs will be taken
# as a baseline (hadron h1). The remaining
# PDFs (starting from the second LHAID in 
# the list) will be used as a PDF for the
# hadron h2.
#
# For details see: arXiv:XXXX
#  
# Works only with: 
#
#    lhapdf = pdlabel
#
#********************************************
\end{verbatim}
To trigger the asymmetric computation one should select:
\begin{verbatim}
#********************************************
 True = asymm_choice 
 ! choice of the collision system
#********************************************
\end{verbatim}
and provide a list of at least two {\tt LHAIDs}.
In the case of $pA$ collisions, the first {\tt LHAID} has to be the one of the proton PDF baseline, while the remaining one(s) correspond to one or more nuclei, e.g.:
\begin{verbatim}
lhapdf = pdlabel
102000, 102400, 103100 = lhaid 
\end{verbatim}
In the example above, the corresponding ids for pro\-ton (102000), Ne (102400) and Pb (103100) PDFs from the nCTEQ15~\cite{Kovarik:2015cma} PDF set are used. The central value for the first PDF in the list is always used as the baseline. The asymmetric \MG code always computes both symmetric and asymmetric quantities, specifically, in the example above it will compute the cross section for both symmetric and asymmetric collisions, {\it i.e.}~$pp$, NeNe, PbPb, $p$Ne, Ne$p$, $p$Pb, Pb$p$ in a single run. 

In the case of a $\pi p$ collision, if one is interested in generating the pion PDF uncertainty, the pion PDF {\tt LHAID} must be the second in the list of {\tt LHAPDF} ids.

As detailed in Sec.~\ref{sec:asym-coll}, the corresponding PDF and/or scale variations are computed when the {\tt run\_card} par\-a\-me\-ters {\tt re\-weight\_PDF} and/or {\tt re\-weight\_scale} are set to {\tt True}. Moreover, we recall that $\alpha_s$ will be taken from the $n-1$ {\tt LHAPDF} sets, starting from the second in the list of {\tt LHAIDs}. Through these simple steps, cross sections and nuclear modification factors with corresponding uncertainties are generated. More specifically, the NMFs are computed directly through the {\tt gnuplot} scripts as ratios of columns of the {\tt HwU} files (containing information about cross sections), and are always available.

Finally, we also offer the possibility to change in which frame the results are computed and plotted, {\it i.e.}~the laboratory frame or the hadronic c.m.s.:

\begin{verbatim}
#********************************************
 False = cms_frame ! choice of the frame 
    ! in which the computation will be 
    ! performed. 
    ! False =  default, laboratory frame, 
    ! True  =  hadron_cms frame
#********************************************
\end{verbatim}
%
Note that this flag is by default hidden in the {\tt run\_card.dat} file and set to {\tt False} such that the colliding particle energies and the kinematical cuts should be entered in the laboratory frame and the output is also given in this frame.
The flag acts during all stages of the run enforcing the kinematic cuts at the level of the event generation, and filling the booked histograms in the selected frame.

\bibliographystyle{utphys}
\bibliography{asym-MG5.bib}

\end{document}